# Ultra-Efficient On-Chip Supercontinuum Generation from Sign-Alternating-Dispersion Waveguides

*Haider Zia[1*], Kaixuan Ye[2], Yvan Klaver[2], David Marpaung[2] and Klaus-Jochen Boller[1]*

Haider Zia, Klaus-Jochen Boller

University of Twente, Department Science & Technology, Laser Physics and Nonlinear Optics Group, MESA+ Research Institute for Nanotechnology, Enschede 7500 AE, The Netherlands

E-mail: h.zia@utwente.nl

Kaixuan Ye, Yvan Klaver, David Marpaung

University of Twente, Department Science & Technology, Nonlinear Nanophotonics Group, MESA+ Research Institute for Nanotechnology, Enschede 7500 AE, The Netherlands

**Abstract:**

Fully integrated supercontinuum sources on-chip are critical to enabling applications such as portable and mechanically-stable medical imaging devices, chemical sensing and LiDAR. However, the low-efficiency of current supercontinuum generation schemes prevent full on-chip integration. In this letter, we present a scheme where the input energy requirements for integrated supercontinuum generation is drastically lowered by orders of magnitude, for bandwidth generation of the order of 500 to 1000 nm. Through sign-alternating the dispersion in a CMOS compatible silicon nitride waveguide, we achieve an efficiency enhancement by factors reaching 2800. We show that the pulse energy requirement for large bandwidth supercontinuum generation at high spectral power (e.g., 1/e level) is lowered from nanojoules to 6 picojoules. The lowered pulse energy requirements enable that chip-integrated laser sources, such as mode-locked heterogeneously or hybrid integrated diode lasers, can be used as a pump source, enabling fully integrated on-chip high-bandwidth supercontinuum sources.

## 1. Introduction

Significantly increasing bandwidth through intensity-dependent nonlinearity, i.e., Supercontinuum generation (SCG) [1–3] is central to a vast plethora of applications such as in frequency combs,[4,5] hyperspectral LiDAR,[6,7] optical coherence tomography (OCT)[8,9] or nonlinear optical pulse compression.[10,11] Most implementations of SCG require significant pump pulse energy to broaden the spectral bandwidth at useful power levels for most applications.[12] Achieving ultra-wide bandwidth spectra remains challenging with on chip pump sources due to the high pulse energy requirements, thereby limiting full integration of laser sources on chip with SCG waveguides.

Currently, chip-scale supercontinuum generation [13,14] requires pump pulse energies in the order of hundreds of picojoules or several nanojoules. However, pumping SCG with chip-integrated laser sources, such as with mode-locked heterogeneously or hybrid integrated diode lasers, [15,16] would require that the supercontinuum generation can be achieved with significantly lower pulse energies in the sub-10 pJ, presenting a major bottleneck in achieving compact SCG sources. The low pulse-energy requirement especially limits bandwidth generation within the high spectral power 1/e (approx. -4 dB) range, crucial for interferometric applications such as OCT and the generation of ultrashort pulses.

Nowadays several approaches have been explored to lower this pump energy requirement. One approach is to utilize materials with a smaller bandgap, which increases a waveguide's effective nonlinear coefficient via an increased nonlinear susceptibility.[17,18] However, this may limit material choices for efficient nonlinear bandwidth generation. As well, employing alternative materials may be hard to implement in CMOS compatible platforms; limiting any widespread practical implementation. Another method relies on enhancing the peak intensity by tightly confining the optical radiation within the waveguide. Thus, choosing a high-index contrast waveguide CMOS compatible platform, such as silicon nitride circuits, further increases the peak intensity and bandwidth. [19,20] These methods eventually fall through because the pulse's peak power is lost or rendered ineffective by the interaction of waveguide dispersion and self-phase modulation. Due to dispersion, either temporal broadening will occur, reducing peak power, or the pulse forms into stationary solitons, both leading to a stagnation of spectral generation above the -20 dB range.[21] Thus, simply increasing the confinement or increasing the waveguide length is not effective in matching the low pulse energy requirements of fully integrated SCG sources.

A promising approach is to alternate the sign of dispersion in segments of different lengths across the propagation to counteract both types of SCG stagnation mechanisms of spectral generation. [22,23] The bandwidth increase, notably at the 1/e level, is then on-going with increasing number of alternating dispersion segments, resulting in a substantially increased spectral generation length, terminating only if waveguide losses start to dominate.

In this letter we report novel implementations to increase spectral generation efficiency of the bandwidth at high spectral power, needed for applications such as OCT or nonlinear pulse compression in integrated waveguide SCG.[24] We experimentally demonstrate SCG generation in repeatedly sign-alternated-dispersion silicon nitride waveguides, that exhibits 1/e bandwidth generation enhancement up to 2800 times compared to the state-of-the-art. The increased energy efficiency of our approach leads to a required pulse energy down to the sub-ten picojoule level. Our results represent an important step towards all-integrated wide-bandwidth portable light sources.

## 2. Results

### 2.1 Design of the Alternating Dispersion Waveguides

A conceptual illustration of the alternating waveguides is shown in **Figure 1a**. In our waveguide design of fixed thickness, stepwise variation of the waveguide width along the propagation direction is used to set the dispersion at normal (ND) or anomalous (AD).

Due to alternating the dispersion, the loss of peak power from temporal broadening in normal dispersion is offset by pulse compression in anomalous dispersion segments. In turn, the cessation of spectral generation due to soliton formation[25] in anomalous segments is overcome by spectral generation in the ND segments and by nonlinear compression of the chirped stretched pulse in the subsequent AD segments. The spectral generation then remains across a longer interaction length of the waveguide resulting in large spectral enhancements of SCG by sign-alternation of dispersion.

To minimize losses, adiabatic tapers are placed to connect the segments in a chain. The pulse compression to shorter durations per subsequent anomalous dispersion (AD) segment and the increasing frequency chirp and duration after the normal dispersion (ND) segments are indicated in the figure by the colored example profiles of the pulse's electric field profile after some of the segments.

As shown in Figure 1a (not to scale), two alternating structures are considered to maximize spectral bandwidth to pump-power efficiency for either the vertical or horizontal input linear polarization state. Structure 1 maximizes the spectral generation of the mode with vertical polarization (**s**), while Structure 2 maximizes that of the horizontal polarization (**p**). Structure 1 has more width iterations than Structure 2.

The two structures were designed using a SCG solver,[23] based on the non-linear Schroedinger Equation, described in Methods**.** The structures were fabricated at the Ligentec foundry with losses of less than 0.05 dB/cm by use of SiN/SiO high-resolution All-Nitride (AN) technology[26] (see Methods and Supplementary Information I on the structures, solver and losses).

The calculated dispersion spectra of the AD and ND segments are compared in **Figure 1b and c** for both the s and p-polarizations. The shaded areas indicate the bandwidth across which the sign of the dispersion is inverted, which is approx. 750 nm for the s-polarization, covering a wavelength range from 1200 nm to 1950 nm. The inversion bandwidth for the p-polarization is 300 nm with wavelength range from 1300 nm to 1600 nm. The expectation is that the generated bandwidth spans at least the sign-inversion wavelength range, where dispersive loss of peak intensity and soliton formation are prevented.

The experimental setup is shown in **Figure 1d** and described in Methods. An input pulse centered at l554 nm, with 21 nm bandwidth and approximate duration of 165 fs was used from an ultrafast fiber laser (Toptica FemtoFiber Ultra). Pulses were coupled into the waveguides through a lensed fiber (pulse duration after lensed fiber is 179 fs, with negligible nonlinearity- shown in methods).

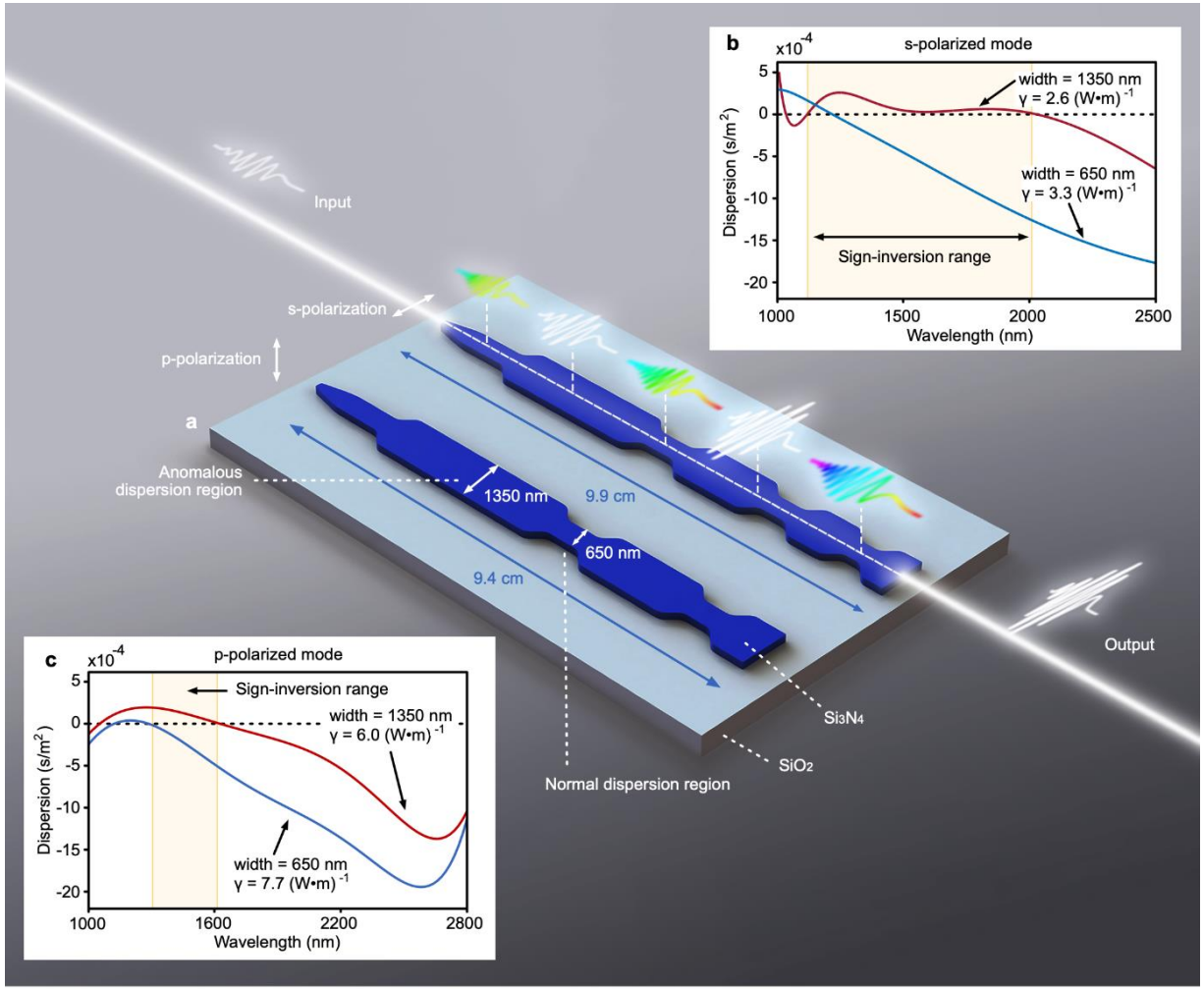

**Figure 1**. **Sign-alternated-dispersion silicon nitride waveguides for efficient supercontinuum generation.** a) An illustration of integrated waveguide Structure 1 and Structure 2. The darker region indicates the silicon nitride core, while the lighter region corresponds to the silicon oxide cladding. The overall lengths of both structures are indicated, and the amount of ND and AD segments are also shown. Nonlinear pulse propagation is shown through the qualitative dynamics of the illustrated electric field profiles. The length and width of the segments are not shown to scale. For the exact geometrical dimensions of the waveguides, refer to Supplementary Information I. b) Calculated dispersion profiles of the s-polarized mode versus wavelength for the AD segment of 1350 nm width (shown in red) and the ND segment of 650 nm width (shown in blue). The associated nonlinear coefficients of the waveguides are indicated as well. The shaded region is the wavelength range where the sign of dispersion is inverted. c) Dispersion profiles for p-polarizations, analogous to panel b). d) Illustration of the experimental setup. A fiber laser delivering ultrashort pulses was used as the source. PBS refers to polarization beam splitter, DM refers to dielectric mirror (1550 nm), SM to broadband silver mirror MM refers to multi-mode fiber. OSA refers to near-infrared or visible range optical spectrum analyzers (Ocean View NirQuest256 and Ando AQ6315A spectrum analyzer).

## 2.2 Low Pulse Energy Spectral Output of the Alternating Dispersion Waveguides

The output power spectra were measured at different pulse energies and polarization states for each waveguide structure. The overall transmission loss trough the waveguide circuits was calculated using the loss values provided by the manufacturer as given in the Supplementary Information I. To determine the in-coupled power independent of uncertainties in mode matching efficiencies at the input, the power in the waveguides was measured at the output and corrected for propagation loss. Using manufacturer provided values and simulations we estimate that losses due to propagation and out-coupling do not exceed a maximum total of 30%. In order to effectively compare the simulations with experiment, we refer to the estimated pulse energy just after in-coupling to the waveguide.

**Figure 2a** shows the output spectral power plotted against wavelength for Structure 1 for the s-polarized input. With increasing pulse energy, both the bandwidth increases and the amount of spectral power approaching or exceeding the 1/e level. However, of main importance is that the structures provide an unprecedented bandwidth generated at low pulse energy. The high bandwidth efficiency reveals itself by the low pulse energy threshold, needed for the 1/e width to exceed the sign-inversion dispersion bandwidth (marked in yellow in Figure 2a) at 34 pJ ( 24 pJ measured after the output lens).

At the threshold pulse energy at 34 pJ, the bandwidth spans from approx. 1500nm to 2450nm, i.e., 950 nm. At the -30 dB level, the spectrum spans more than an octave. As pulse energy increases, the spectrum flattens within the 1/e width which in turn increases to greater than 1000 nm at an output pulse energy of 55 pJ. Increasing pulse energy also results in narrow band dispersive wave components at approx. 480 nm, 515 nm and 550 nm, shown in the inset.

An example spectrum obtained from our solver is shown as the dotted curve in Figure 2a. While the simulated spectrum has a pulse energy of 20 pJ, it best matches the experiment at pulse energy of 34 pJ in terms of profile and bandwidth. The offset is likely from remaining deviations between the assumed numerical dispersion profiles and nonlinear coefficients and the actual profiles obtained from the fabrication process. However, while these deviations are present, as can be seen from the experiments, the simulation was able to still find an efficient sign-alternating dispersion structure that maximized the 1/e bandwidth range, describes the salient dispersive and nonlinear effects and shows the expected shape of the spectral profile.

As shown in **Figure 2a and b**, it was experimentally found that for both structures, the spectral bandwidth exceeds the sign-inversion wavelength range. We first explain this finding, to supplement the above discussion on the nonlinear pulse dynamics in sign-alternating dispersion structures.

As confirmed by simulations, the efficient spectral bandwidth generated outside the inversion region is from dispersive waves generated in the AD and ND segments as follows. As the optical pulse repeatedly undergoes spectral generation in subsequent AD and ND segments, dispersive waves are also repeatedly generated. However, the main pulse profile changes (e.g., in peak intensity) due to the nonlinear propagation within each segment, generating a different frequency profile of dispersive waves at each segment. The superposition of all the generated dispersive waves then creates the broadband spectral profile above the 1/e-level and outside the inversion range, seen in the output spectra (refer to Supplementary Information II for more details).

In the presentation of spectra in this paper, we start at the pulse energy when the formation of the broadband dispersive waves surpasses the 1/e level. However, substantial spectral broadening (e.g., >200 nm) occurs already at lower pulse energies, but with a 1/e-level bandwidth contained within the inversion wavelength range.

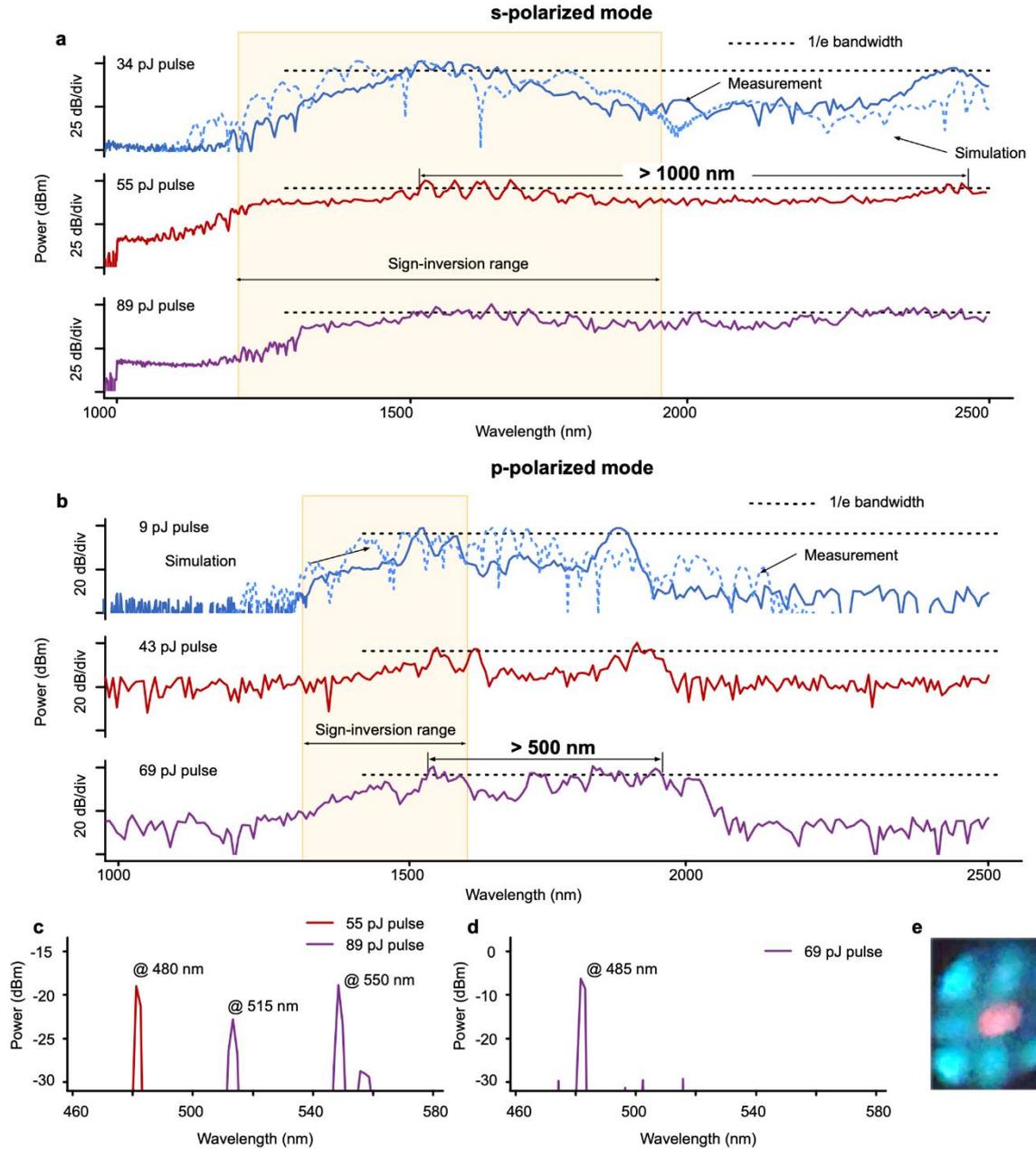

**Figure 2. Measured supercontinuum spectra from the silicon nitride waveguides**. a) Spectral power versus wavelength on a logarithmic scale for Structure 1's s-polarization mode, recorded at stepwise increased pulse energies as indicated in plots. Dispersive waves in the visible range are indicated in inset. b) Same plot as for a) for Structure 2's p-polarization mode. c) spectral power of the visible dispersive waves of Structure 1 at a pulse energy of 55 pJ and 89 pJ. d) spectral power of the visible dispersive waves of Structure 2 at a pulse energy of 69 pJ. e) Real-color image of an example of higher order modes of the visible dispersive wave at 485 nm. The pink coloring in the center is due to the IR sensitivity of the camera.

The total bandwidth generated is generally larger with wider inversion bandwidth, as seen when comparing the s-polarization bandwidth to that of the p-polarization for Structure 1 and 2 respectively (Figure 2b). For Structure 2, designed to maximize the bandwidth of the p-polarization, a 1/e spectral width of approx. 375 nm is obtained at 9 pJ pulse. The simulation trace is shown as the dotted blue line. Narrowband high-power visible dispersive waves, exceeding the -20 dB level are found between 480 nm and 580 nm in Structure 1 (**Figure 2c**). For structure 2 a dispersive wave at 485 nm exceeds the -8 dB level as shown in **(Figure 2d)**. The visible dispersive waves for Structure 1 and 2 exhibit higher-order spatial modes. For example, the dispersive wave at 485 nm is approximately a TE$_{33}$ mode as shown in **Figure 2e**.

The visible dispersive waves are in higher order modes and thus, we surmise that they are not from the traditional SPM-GVD dynamics but result from spontaneous four-wave mixing in the AD segments (i.e., they are close to the third-harmonic) and then are amplified by the increased spatial overlap of the radiation in the more confined ND segments.

Additional spectra for the p-polarization for Structure 1 and s-polarization for Structure 2 are shown in Supplementary Information III. The bandwidth and spectral profiles for the other polarizations for Structure 1 and 2 are similar to the corresponding polarizations presented here, albeit occurring at higher output pulse energies.

## 3. Discussion

To put our results in context, we have defined an efficiency enhancement factor ($\eta_{BP}$) to compare the spectral efficiency versus input pulse energy of our sign-alternated devices with other integrated SiN SCG setups. The efficiency enhancement factor gives the gain of efficiency of our work compared to others and is given as:

$$\eta_{BP} = \frac{\Delta\lambda_{alt}}{\Delta\lambda_{ref}} \frac{E_{ref}}{E_{alt}}. \quad (1)$$

In **Equation 1**, $\Delta\lambda$ refers to the output 1/e bandwidth and 'E' refers to input pulse energy. The index 'alt' represents alternating dispersion for our p and s-polarization structures while 'ref' indicates previously reported experimental data without alternation.

Equation 1 provides a means of quantifying various SCG schemes across a wide parameter range. The $\eta_{BP}$ as defined should be seen with some caution as it cannot incorporate all the specifics of each case, and is therefore only a global measure. On the other hand, it is independent of specific waveguide parameters, only looking at the overall efficiency of state-of-the art systems as compared to ours.

We tabulate the efficiency enhancement factor of our method in comparison to several state-of-the-art experimental results for SCG in the integrated silicon nitride platform[12,13,27–32], shown in Table 1 for both structure 1 (s-polarization input) and structure 2 (p-polarization input). With regard to all of the compared works, the alternating SCG waveguides show a clear relative advantage for both polarizations, and a maximum efficiency enhancement factor of $\eta_{BP} > 2700$.

**Table 1. Efficiency comparison to previous state-of-the-art integrated SCG sources.**

| Technology | Materials | Pump Pulse Input | | SCG 1/e Bandwidth (nm) | Efficiency (nm/pJ) | Efficiency ratio of our work to cited work | |
| --- | --- | --- | --- | --- | --- | --- | --- |
| | | FWHM (fs) | Energy (pJ) | | | | |
| This work: s-pol. Optimized alternating waveguide | silicon nitride on oxide | 180 | 34 | 950 | 28 | | |
| This work: p-pol. Optimized alternating waveguide | silicon nitride on oxide | 180 | 9 | 375 | 42 | s-pol. Optimized alternating waveguide | p-pol. Optimized alternating waveguide |
| two segment - either width or cladding removal - cascaded anomalous dispersion waveguide[33] | silicon nitride on oxide, 100 mm length | 60 | 220 | 83 | 0.4 | 74 | 110 |
| uniform-width anomalous waveguide [32] | silicon nitride on oxide, pumped at 1030 nm | 115 | 590 | 150 | 0.3 | 110 | 160 |
| wide (1300 nm) uniform-width anomalous waveguide[13] | silicon nitride on oxide | 120 | 1400 | 21 | 0.02 | 1900 | 2800 |
| Bragg grating enhanced SCG[27]‡ | silicon rich nitride on oxide (USRN) | 1700 | 10 | 24 | 2.4 | 12 | 17 |
| enhanced confinement and | silicon on insulator (SOI) | 250 | 27 | 400 | 15 | 2 | 3 |
| tapered anomalous waveguides[28]* | pumped at 1950 nm | | | | | | |
| Bragg grating and confinement enhanced SCG[29]* | silicon on insulator (SOI) | 250 | 27 | 350 | 13 | 2 | 3 |
| previous state of the art in efficiency SCG OCT source (operating at 1300 nm) [12] | silicon nitride on oxide pumped at 1300 nm | 200 | 25 | 110 | 4.4 | 6 | 9 |
| uncladded high confinement anomalous dispersion waveguide[30] | silicon nitride on oxide | 78 | 300 | 1000 | 3.3 | 8 | 13 |
| cross phase modulation intermodal SCG[31] | silicon nitride on oxide | 88 | 250 | 100 | 0.4 | 70 | 100 |
| Low pulse energy SCG [34] | Silicon nitride pumped at 1064 nm | 70 | 18 | 5 | 0.3 | 101 | 150 |

The asterixis refer to works with a waveguide nonlinear coefficient approx. a factor of around 3 times higher than what was used in our structures. The double cross refers to a work in silicon rich nitride (USRN) with a nonlinear refractive index and coefficient approx. an order of magnitude higher than our structures.

To signify the wide parameter range encompassed by the references shown in Table 1, we also include cases where alternating dispersion does not seem to offer an advantage on a first view, as the efficiency enhancement factor is comparatively low ($\eta_{BP} \approx 3$ and 12, marked by asterixis or double-crosses in Table 1). However, closer inspection shows that also here there is a clear advantage, because these works used waveguides with a nonlinear refractive index and coefficient an order of magnitude higher, e.g., with silicon rich nitride (USRN – indicated by double-crosses), or that have a nonlinear coefficients around 3-times higher (indicated by asterixis).

Most closely matching our own pulse duration (180 fs) are the most recent experiments described in Ref. 12 (200 fs). In that example a state-of-the-art integrated OCT source is demonstrated, showing a spectral width of 100 nm at 25 pJ, while our approach provides a width of 950 nm using 34 pJ pulses, yielding a relative efficiency enhancement factor of $\eta_{BP} = 9$, i.e., input pulse energy requirements are lowered by almost an order of magnitude.

The structures in Table 1 span a length shorter than our structures. In our approach, the AD segments lengths are lower than the soliton fission length but are terminated right before soliton fission occurs. This condition ensures that spectral generation in the AD segments is maximized and that solitons would not be generated. We avoid that solitons are generated because any spectral generation is then limited but modulations in the spectrum increase, i.e., the spectral quality degrades. In fact, past soliton fission, the spectral bandwidth may even decrease in the AD segments or within the ND segments because of the temporal distribution of the soliton bunch center frequencies, i.e., they have an opposite chirp profile to what generates spectrum in the SPM-GVD dynamics.[35]

Since the soliton fission length increases at lower pulse energy, the footprint of our structures, spanning approx. 9 cm on a 5 mm by 10 mm chip is larger than structures considered above in the table. However, the longer AD segments to utilize the full range possible for spectral generation, coupled with sign-alternation is what generates the high efficiencies recorded in Table 1 of our approach, and enables large 1/e bandwidth generation at low pulse energy (see supplementary information for more details).

**4. Conclusion**

In conclusion, alternating dispersion waveguides for wide-bandwidth generation at the 1/e bandwidth are an important step forward for all-integrated SCG sources. Our approach can be implemented in a variety of material platforms where dispersion can be engineered by the geometry of the waveguide, e.g., in high contrast refractive index materials.

In regards to past work in dispersion management for SCG, the use of complex machine learning algorithms has been explored recently in Silicon on Insulator where 1/e bandwidths of 300 nm has been achieved at 4 pJ but through a high nonlinear coefficient of 220 $(Wm)^{-1}$.[36] However, our approach indicates that more efficient SCG 1/e bandwidth generation can take place by simply alternating the sign of dispersion , in materials with low nonlinearity (at most 8 $(Wm)^{-1}$). It would be of interest to compare the findings of the machine learning algorithm to our approach of sign-alternation in a waveguide system with the same material nonlinearity for both methods. If sign-alternation produces comparable or better efficiency than the reduction in complexity can narrow the parameter space for SCG machine learning algorithms.

The pulse energy needed for wide-bandwidth SCG at the 1/e width, found to be 9 pJ at 1550 nm in our approach, is now coming into proximity with the 1-3 pJ pulse range offered by state-of-the-art integrated ultrafast diode lasers centered at 1550 nm. [15,16] Ultra-short pulse durations are supported by the bandwidth of the laser described in Ref. 16, and the required moderate upscaling of the pulse energy appears feasible, e.g., with on-chip amplification.[37]

Diode pumped solid state waveguide lasers are more challenging regarding CMOS compatibility although the required optical pumping has shown great progress with off-chip pump lasers. While these lasers operate so far only in the CW regime[38] they hold potential as a laser source for the sign-alternating SCG waveguides presented in this letter, since noticeably higher upper-state lifetimes and, correspondingly, higher pulse energies are offered.

## 5. Experimental Section/Methods

### 5.1 Experimental Setup

A Toptica FemtoFiber Ultra laser was used, centered at 1554 nm, with 78 MHz repetition rate and a pulse duration of 165 fs full-width half maximum. The laser is operated at maximum power, so the output pulses have the widest spectral profile. The experimental setup is shown through the schematic in Figure 1 d), in the main text.

The source pulses passed through an isolator to minimize feedback. A half-wave plate and polarization beam splitter (PBS) arrangement was then placed to serve as a variable power attenuator.

After the average power is lowered, the laser pulses are passed through another half-wave plate and quarter-wave plate arrangements, such that the polarization state is set to the s or p polarization entering the alternating waveguide downstream. The quarter-wave plate is necessary to account for the birefringence in the lensed fiber used to couple light into the waveguide.

Next, the pulses pass through a lensed fiber (OZ optics, TSMJ-3S-1550-9/125-0.25-5-5-45-0.28-POL-AR) consisting of SMF28 fiber of approx. 30 cm in length, terminated by a taper down to a tip diameter of 5 μm. The lensed fiber maximized the power coupled into the waveguide (21%), calculated from the output power and estimated propagation loss. The dispersive effects were then verified to be negligible by measuring the pulse duration at the output of the fiber, using an autocorrelator (APE PulseCheck NX NIR), where the autocorrelation trace is shown in Figure 3a. The associated Gaussian pulse duration at the FWHM is approx. 179 fs which is similar to that of the source laser (164 fs). Nonlinear effects in the fiber were verified to be negligible by measuring the spectrum of the pulses at the fiber end-facet using an Ando AQ6315A spectrum analyzer (Figure 3b). Figure 3b shows the spectral power after the lensed fiber for the maximum power used into the chip compared to the spectral power before the lensed fiber.

a)
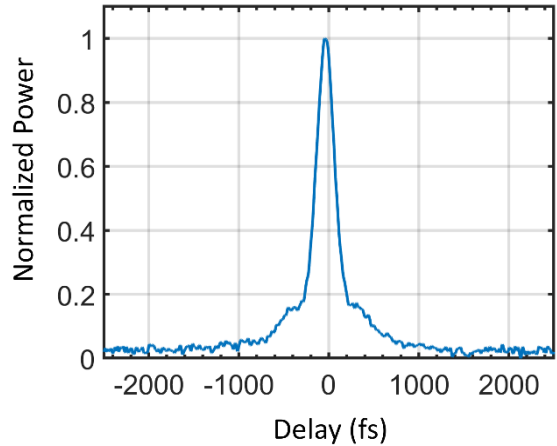

b)
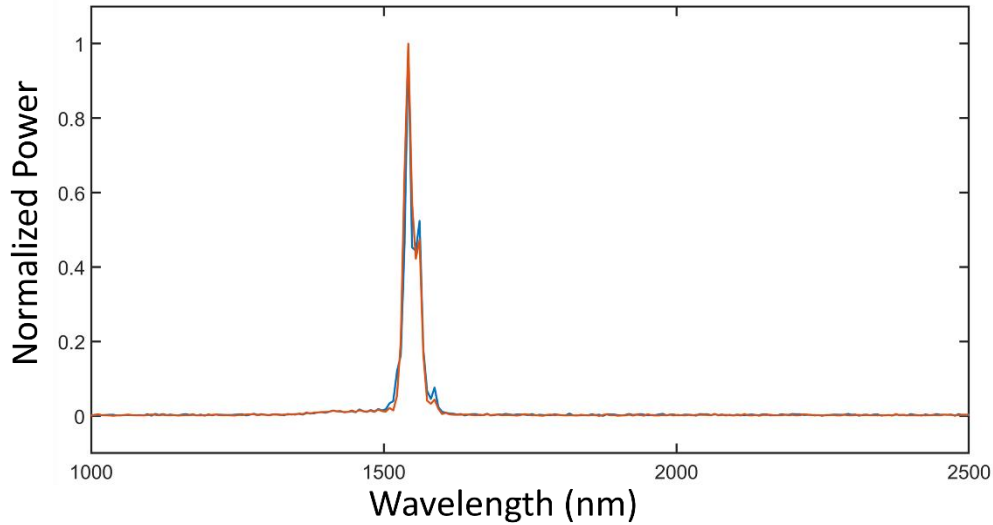

Figure 3: a) Autocorrelation trace at maximum power used for the experiment after the lensed fiber. b) Spectrum at powers used in the experiment, after the lensed fiber (blue), compared to spectrum inputted into lensed fiber (red).

The fiber is mounted on a 5-axis stage (Nanomax Thorlabs, 3 translational, pitch and yaw) to maximize input coupling. The waveguides are mounted on a separate stage with only translation perpendicular to the propagation axis. The pulses were out-coupled from the waveguides and collimated using a 3.1 mm aspheric focus lens (Thorlabs C330TMD-C) on a three-axis stage (Nanomax, Thorlabs). The near-infrared spectrum was measured using a NIR spectrometer ( Ocean View NirQuest256 ) with range from 850 nm to 2500 nm. The portion of the spectrum, from 350 nm to 850 nm was measured with the Ando spectrum analyzer whose total range extends from 350 nm to 1750 nm.

## 5.2 Waveguide Design and Fabrication

The waveguide width profile versus propagation coordinate was found, by solving the associated group-velocity dispersion profile using a numerical solver (Supplementary Information I) and this information was the basis for the fabrication of the structure. Parameters for the waveguide, such as the inverse input taper and the adiabatic taper lengths between segments, losses and minimal bend radii, were provided by VLC photonics. The chip-layout and wafer run were designed and coordinated by VLC Photonics as well.

The fixed thickness of our waveguides is 800 nm, and the width of the ND segments is 650 nm while that of the AD segments is 1350 nm. Adiabatic tapers, with a length of 59 um, are placed to connect the segments. Structure 1, has a total length of 9.9 cm, and Structure 2 has a length of 9.4 cm. Both have an input inverse taper starting at a core width of 200 nm. Both structures are folded on a 2-cm by 1-cm chip where the bend radii are chosen such that the bends negligibly influence the optical radiation.

## 5.3 Theory of Nonlinear Pulse Propagation and Supercontinuum Generation

The experiment was modeled using a generalized nonlinear Schroedinger equation (GNLSE) under the slowly varying envelope approximation, given in **Equation 2**:[17]

$$\frac{\partial u}{\partial z} = \sum_{k \geq 2} \frac{i^{k+1}}{k!} \beta_k \frac{\partial^k u}{\partial \tau^k} + i\gamma (1 + i\tau_s \frac{\partial}{\partial \tau}) |u|^2 u \quad , \tag{2}$$

where u is the complex field envelope and $\beta_k$ are the Taylor series coefficients of the expansion of the frequency-dependent wave number about the central frequency, $\omega_o$. Since the expansion starts at $k = 2$ this amounts to the Taylor series expansion of the group-velocity dispersion. $\tau = t - V_g z$ is the time coordinate, comoving in the frame of reference of the group velocity dispersion ($V_g \equiv \beta_1^{-1}$). $\tau_s$ is the characteristic timescale of self-steepening (shock time), given as approx. 5 fs. The equation was evaluated using the split-step exponential Fourier method[39] iteratively along the propagation coordinate z. The procedure to obtain the frequency-dependent group velocity dispersion of the segments is described in Supplementary Information I.

**Supporting Information**

Supporting Information is available from the Wiley Online Library or from the author.

**Acknowledgements**

The authors gratefully acknowledge Toptica Photonics AG for the pulsed laser source, Ocean Insight for the NIR spectrometer and APE GmBH for the NIR autocorrelator. The authors also acknowledge VLC Photonics and Ligentec SA for the silicon nitride chip fabrication.

**Funding**

Haider Zia would like to acknowledge funding from the NWO through the Demonstrator Grant, project number 18562.